\documentclass[sigconf,10pt]{acmart}
\usepackage{cleveref}
\usepackage{arydshln}
\usepackage{subcaption}
\makeatletter
\renewcommand\paragraph{\@startsection{paragraph}{4}{\z@}%
  {1ex \@plus0ex \@minus0ex}%
  {-2ex}%
  {\normalfont\normalsize\bfseries}}
\makeatother

\newcommand{\hspec}[1]{\textcolor{blue}{#1}}
\newcommand{\lspec}[1]{\textcolor{red}{#1}}
\newcommand{\uGPU}{Lite-GPU}
\newcommand{\uShort}{Lite}

\title{Good things come in small packages: Should we build AI clusters with {\uGPU}s?}

\author{\rm {Burcu Canakci*, Junyi Liu*, Xingbo Wu*, Nathanaël Cheriere, Paolo Costa, Sergey~Legtchenko, Dushyanth Narayanan, Antony Rowstron}}
\affiliation{%
  \institution{Microsoft Research}%
  \city{}
  \country{}%
}
\thanks{{\textasteriskcentered}equal contribution, \{burcucanakci,junyili,xingbowu\}@microsoft.com}

\renewcommand{\shortauthors}{Burcu Canakci, Junyi Liu, Xingbo Wu et al.}

\begin{abstract}

To match the blooming demand of generative AI workloads, GPU designers have so far been trying to pack
more and more compute and memory into single complex and expensive packages.
However, there is growing uncertainty about the scalability of individual GPUs and thus AI clusters,
as state-of-the-art GPUs are already displaying packaging, yield, and cooling limitations.
We propose to rethink the design and scaling of AI clusters
through efficiently-connected large clusters of \emph{\uGPU}s, GPUs with single, small dies and a
fraction of the capabilities of larger GPUs.
We think recent advances in \emph{co-packaged optics} can enable distributing AI workloads onto many {\uGPU}s through high bandwidth and efficient communication.
In this paper, we present the key benefits of {\uGPU}s on manufacturing cost, blast radius, yield, and power efficiency;
and discuss systems opportunities and challenges around resource, workload, memory, and network management.

\end{abstract}

\copyrightyear{2025} 
\acmYear{2025} 
\makeatletter
\def\@ACM@copyright@check@cc{}
\makeatother
\setcopyright{cc}
\setcctype{by-nc-nd}
\acmConference[HOTOS '25]{Workshop on Hot Topics in Operating Systems}{May 14--16, 2025}{Banff, AB, Canada}
\acmBooktitle{Workshop on Hot Topics in Operating Systems (HOTOS '25), May 14--16, 2025, Banff, AB, Canada}
\acmDOI{10.1145/3713082.3730390}
\acmISBN{979-8-4007-1475-7/2025/05}

\begin{document}

\maketitle

\section{Introduction}

\begin{figure}[t]
    \centering
    \includegraphics[width=\columnwidth]{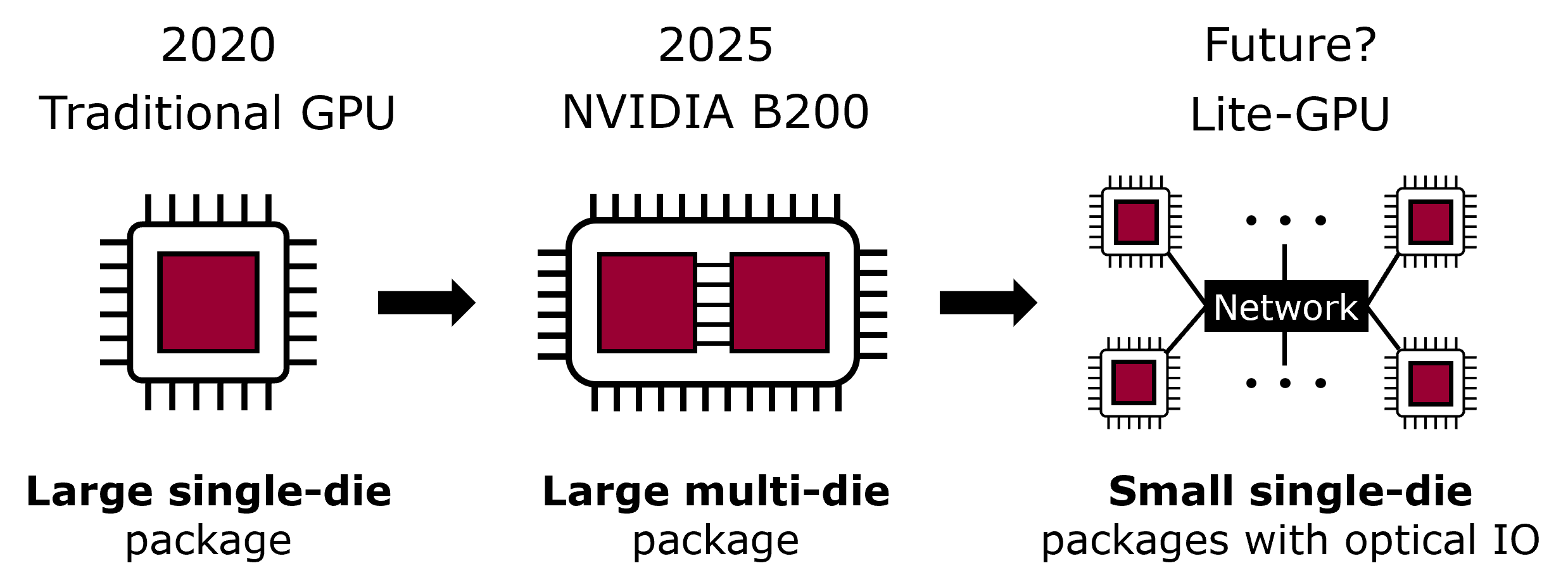}
    \caption{\textmd{Evolution of GPUs in AI clusters.}}
    \label{fig:evol}
\end{figure}

Demand for AI is growing and expensive to support~\cite{modelsurvey}.
These challenges are expected to only get harder as the diversity, complexity, and scale of AI models are growing,
making it crucial for AI service providers to build powerful and efficient AI infrastructure~\cite{openai_infra}.
However, scaling AI infrastructure is encountering significant obstacles~\cite{blackwelldelayed}.
We have already reached the limit on how big a compute die can get,
leading GPU designers to focus on advanced packaging technologies to pack more transistors into the same package (\Cref{fig:evol}). 
Nevertheless, scaling an individual GPU package is becoming less and less sustainable for manufacturing due to multiple reasons,
including power~\cite{blackwell}, cooling~\cite{liquid1}, yield~\cite{yield1,yield2}, packaging costs~\cite{packagingcost}, and failure blast radius~\cite{failure1}.
For instance, the latest generation of NVIDIA GPUs is facing deployment delays due to packaging and cooling issues~\cite{blackwelloverheat, cowos}.

We observe that there is an exciting alternative approach to scaling AI clusters. 
\textit{What if we replace large, powerful GPU packages with highly-connected clusters of {\uGPU}s, 
that each have only a single, smaller compute die and fractional performance?} 
Smaller GPUs present many promising hardware characteristics: they have much lower cost for fabrication and packaging, 
higher bandwidth to compute ratios, lower power density, and lighter cooling requirements. 
In addition, they can also unlock desirable systems opportunities such as improved fault-tolerance and finer-grained, flexible resource allocation.

To date, distributing AI workloads to large number of GPUs has been challenging due to the data flow demanding very 
high-bandwidth communication across GPUs~\cite{paralleltraining}. 
Nevertheless, driven by recent advances in \emph{co-packaged optics}, in the next decade,
we expect off-package communication bandwidth to improve by 1--2 orders of magnitude 
with much better reach (10s of meters), compared to copper-based communication~\cite{cpo1,cpo2, opticalreach}.
Co-packaged optics integrates electronic and optical components within millimeters, compared to 
current pluggable optics, cutting signalling distance and yielding better power efficiency.
While there are open questions and active research on utilizing co-packaged optics, 
we think that it has the potential to disrupt the trade-off space around designing AI infrastructure. Notably, in the most recent GPU Technology Conference (GTC), NVIDIA highlighted their advancements in co-packaged optics in their effort to massively scale AI infrastructure with much improved power efficiency~\cite{nvidia_cpo_gtc}. We think co-packaged optics can enable {\uGPU}s that are equipped with high-bandwidth and energy-efficient optical interconnects, to communicate with many far-off {\uGPU}s at petabit per second bandwidths~\cite{cpo1,cpo2}.

In this paper, we look at AI infrastructure through the lens of {\uGPU}s.
Though we give an overview of recent hardware trends and of the key hardware benefits of {\uGPU}s, 
we focus mainly on the systems opportunities and challenges that would arise as we include {\uGPU}s in AI infrastructure.
We discuss how {\uGPU}s could be beneficial in improving resource customization, resource utilization, power management,
performance efficiency, and failure blast radii in an AI cluster.
In addition, as an initial assessment, we do a performance analysis of a {\uGPU} cluster using popular large language model (LLM) inference workloads.
We show that {\uGPU}s have the potential to match or achieve better performance compared to existing GPUs, as they 
exploit hardware potentials offered by increased total shoreline bandwidth per compute and reduced power density.
These benefits can not be realized for free: we identify key research problems around building a cheap and efficient network,
co-designing the AI software stack, and data-center management.

\section{The {\uGPU}}\label{sec:ugpu}

In recent years, state-of-the-art data-center GPUs have been increasing compute FLOPS, memory bandwidth, and network bandwidth to support growing AI workloads.
As we have already reached the limit of what can be done with a single die~\cite{retical}, improvements have relied on advanced packaging 
efforts to pack more transistors into the same GPU. 
For example, most recently, NVIDIA has featured a multi-die GPU design, using high-bandwidth die-to-die interfaces to bind two dies in its Blackwell GPU platform~\cite{blackwell}. 
Alternatively, AMD has proposed chiplets, breaking up monolithic silicon into smaller specialized chips, co-packaged together through 3D stacking~\cite{chiplet1}.
While these techniques have succeeded in improving GPU performance for their generation, there is not a clear path to scaling them further, and in fact, such complex GPU designs are already leading to several difficulties such as maintaining high yield rates, managing high power consumption, and applying efficient cooling~\cite{packagingcost, yield1, yield2, liquid1}. Additionally, as the die gets larger, its area increases faster than its perimeter (``shoreline'') that determines the bandwidth it can utilize. This leads to GPUs with high compute-to-bandwidth ratios, which is not always the best fit for AI workloads and results in compute underutilization~\cite{sarathi}. 

\begin{figure}[t!]
    \centering
    \includegraphics[width=\columnwidth]{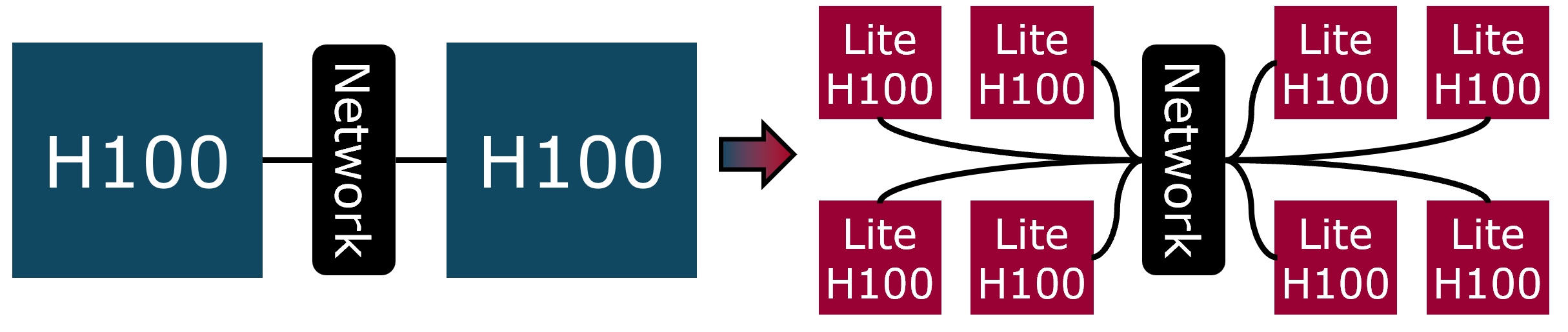}
    \caption{\textmd{An example {\uGPU} deployment. Each NVIDIA H100 GPU is replaced with four {\uGPU}s, featuring better hardware yield and higher bandwidth-to-compute.}}
    \label{fig:ugpu}
\end{figure}

Through {\uGPU}s, we propose an alternative way of scaling AI clusters: with smaller but more GPUs connected through a performant and scalable network, realized through co-packaged optics. A {\uGPU} features a single compute-die GPU package where the die area is much smaller than that of state-of-the-art, leading to several hardware benefits.
\Cref{fig:ugpu} gives an example of a {\uGPU} system where each NVIDIA H100 GPU is replaced with four {\uGPU}s.
In this paper, we mainly use this example while discussing potential benefits of {\uGPU}s in AI clusters.

First, as the die area is smaller per GPU, {\uGPU}s have largely reduced cost of manufacturing due to higher hardware yield rates. For example, the yield rate can be increased by $1.8\times$ when a H100-like compute die area is reduced by $1/4^{th}$, corresponding to almost 50\% reduction in manufacturing cost~\cite{yield_calc}. 

Second, reducing the compute die area increases the shoreline to die ratio. For example, reducing the die area to $1/4^{th}$ doubles the perimeter exposed to the four dies, yielding a cluster with $2\times$ the bandwidth-to-compute ratio.
Although a fraction of the extra bandwidth may be required for additional networking, we show later in our case-study that {\uGPU}s can achieve higher performance efficiency for I/O-bound workloads, such as parts of LLM inference.

Third, smaller packages also greatly reduce complexity of cooling. Today's cutting-edge GPUs already throttle compute frequency to avoid overheating~\cite{thottle_heat, thottle_heat2}. Smaller single-die GPUs can be air-cooled separately and even sustain higher clock frequencies without requiring advanced cooling.

Overall, we expect the cost of {\uGPU}s to be substantially lower due to better hardware yield and lower packaging costs. While the cost of networking should increase, we expect the net gains to be positive as the networking costs are only a small fraction compared to the GPU costs today. Additionally, there are many active efforts to scale networking costs sublinearly with network size using circuit switching~\cite{sirius, tpuv4}, which would allow for even larger {\uGPU} clusters.

\section{Systems opportunities}\label{sec:systems}
\label{sec:sys}

Consider a cluster of NVIDIA H100 GPUs, which is the most frequently deployed GPU in AI clusters today. Each H100 GPU can be replaced with a number of {\uShort}-H100 GPUs, each {\uGPU} having a fraction of its compute and memory capabilities. Depending on how the {\uGPU}s are customized, compared to the original cluster, the cluster with {\uGPU}s can feature equivalent or better compute, memory, and cost characteristics. 

While scaling out AI clusters using today's GPUs with co-packaged optics is an option, as highlighted in the previous section, {\uGPU}s offer many hardware benefits over current GPUs. So in this paper, we focus on utilizing {\uGPU}s as they have potential to unlock the path towards more efficient and scalable AI clusters. Nevertheless, some key systems research questions should be addressed so that we can realize the {\uGPU} disruption.

\paragraph{Scale of distribution} Some of the research questions around using {\uGPU}s are not new or unique, but potentially amplified. For example, {\uGPU}s would result in more distributed systems in the datacenter, e.g., small models previously served by a single GPU are now distributed over multiple {\uGPU}s. For larger models that already require multiple GPUs, the number of devices would be multiplied. These can potentially amplify issues such as synchronization and straggling GPUs. 

AI clusters come at different scales for training and inference, with training clusters being orders-of-magnitude larger, e.g., 16,000 vs 8 GPUs for Llama 3.1 405B~\cite{llama_train, llama_infer}. An inference cluster with {\uGPU}s at the reduction ratios we discussed in \Cref{sec:ugpu} is unlikely to have more components than a training cluster today and would be easier to realize without heavy innovation in distributing models. In general, building efficient distributed ML training and inference platforms is an active research area and such approaches would also benefit clusters with {\uGPU}s ~\cite{deepseekv3, megatron,vllm,deepinf}.

\paragraph{Finer-granularity of resource management} With {\uGPU}s, we can allocate and access smaller units of compute and memory, leading to greater flexibility in managing an AI cluster. 

For example, consider \emph{power management}. A GPU's compute clock frequency can be dynamically tuned to lower power consumption during idle periods or to match stragglers~\cite{perseus, userve}.
However, the granularity of down-clocking is on all \emph{Streaming Multiprocessors} (SMs).
SMs are processors designed for efficient parallel processing and each GPU consists of multiple SMs, similar to cores in a CPU.
Down-clocking all SMs of a large GPU can lead to wasted resources or suboptimal performance.
In a {\uGPU} cluster, we can control down-clocking at finer granularity to achieve better power efficiency, akin to down-clocking only a portion of SMs in a larger GPU.

Conversely, we can over-clock {\uGPU}s to achieve higher performance while serving peak workloads, since smaller die areas allow for easier cooling and higher clock frequencies.
Alternatively, more {\uGPU}s can be utilized to satisfy the peak load, but with the additional power overhead due to increased networking. Detailed analysis on workload patterns and power modelling can help us determine the most power-efficient approach for serving typical and peak workloads with {\uGPU}s.

Another example of resource management is around \emph{GPU configuration}. Note that today, AI clusters with heterogeneous GPUs are already used to serve requests as power-efficiently as possible, e.g., by deploying different phases of transformer inference on different GPU hardware~\cite{splitwise}. We can customize and deploy {\uGPU}s for different profiles of AI workloads, similar to Splitwise, but at much finer scale, e.g., racks of custom {\uGPU}s as opposed to clusters of custom racks. Also, {\uGPU}s can allow for \emph{both} easier over-clocking and higher bandwidth-to-compute ratios, potentially achieving higher performance efficiency at cluster-scale \cite{dvfs_meta,overclock_ms}. 

Third, these smaller GPU units may assist future \emph{AI as a service} offerings. The ability to allocate small customizable {\uGPU} clusters per customer, that are separated physically and provide isolation and security, can be quite powerful. 

\paragraph{Workload management} Careful workload parallelisation, deployment, and scheduling is a must in order to obtain the benefits of {\uGPU}s and to mask their overheads.

Most importantly, with {\uGPU}s, we move previously in-silicon traffic to an optical network, potentially inducing additional \emph{latency} and \emph{network load}. There are workloads that would be challenging to distribute further using {\uGPU}s, such as workloads that introduce randomness and congestion to the network traffic. Nevertheless, with AI workloads, there are several techniques we can use. 

First, AI workloads are highly predictable and pipelined so extra latency can be masked through pre-fetching~\cite{gpuprefetch}. In fact, since {\uGPU}s can feature a higher memory bandwidth-to-compute ratio, they may even allow for reduced request-level latency in AI workloads, as less \emph{batching} may be required to improve compute utilization.

Second, large ML models today are already distributed over many GPUs and communicate through highly efficient collectives to minimize the amount of data exchanged, e.g., through tensor parallelism while calculating matrix-matrix multiplications. One can increase the level of tensor parallelism on a deployment of {\uGPU}s to minimize the end-to-end latency.

\paragraph{Fault-tolerance}
Reducing the size of the GPU naturally reduces the blast radius should a GPU fail due to excessive temperatures, dust or debris, or transistor faults; leading to higher available FLOPS, memory capacity, and memory bandwidth at any time. 

To maximize the benefit from smaller blast radii, building a robust and efficient software stack is crucial. Note that today’s large-scale inference pipelines already impose larger blast radii than the hardware-imposed blast radii: if one GPU out of a group of GPUs serving a model instance fails, the entire instance is taken offline~\cite{tpuv4}. Active work on resolving this issue can also help with {\uGPU} clusters~\cite{faulttol1, faulttol2}. One approach to dealing with such rigid, software-imposed GPU configurations is to include \emph{hot spares}, spare GPUs that can be activated to serve a model instance while recovering from a failure. {\uGPU}s can suit this approach particularly well as a cluster of {\uGPU}s are larger with each additional {\uGPU} being smaller and cheaper. This reduces the proportional overhead of including spare {\uGPU}s, though we still need a strategy for how to best utilize them during normal operation.

In general, {\uGPU}s can help improve fault-tolerance of AI infrastructure. Nevertheless, with {\uGPU}s, the number of GPUs in the cluster is increased and additional networking components may be necessary, potentially leading to different failure frequencies and profiles. A thorough analysis of failures and recovery schemes is necessary to ensure that the reduced blast radius of {\uGPU}s are utilized.

\paragraph{Memory management} Each {\uGPU} has only the fraction of the memory capacity of a larger GPU.
This can be a problem for workloads that require high memory capacity and do not distribute efficiently.
So, there are many open questions about the design of the memory system in a cluster of {\uGPU}s.
For instance, do we need memory-sharing across multiple {\uGPU}s to be an option? What should shared memory semantics look like, e.g.,
do we need to operate with a load/store GPU-to-memory network across {\uGPU}s to prevent extra HBM usage due to network buffering?
Additionally, in a heavily-accessed shared memory setting, how can we alleviate the programming and performance challenges that stem from different tiers of memory?

Another potential approach is to use {\uGPU}s along with disaggregated memory~\cite{disaggmem}. Disaggregated memory can be used to provide a larger memory pool for {\uGPU}s and to allow for more efficient memory sharing across {\uGPU}s, though it introduces additional complexity in memory management. Note that, combined with the finer-granularity of {\uGPU}s, an AI cluster with {\uGPU}s, co-packaged optics, and disaggregated memory can enable us to flexibly adjust the compute-to-memory and compute-to-network ratios per {\uGPU} in the cluster.

\paragraph{Network management} Through {\uGPU}s, communication previously in-silicon in a large GPU is now on the {\uGPU} to {\uGPU} network.

Firstly, the total traffic in a cluster and the total power consumption of the network can be higher. Secondly, in-silicon traffic assumes very high-bandwidth, low latency, and energy-efficient communication. Since the performance and efficiency of communication is degraded outside of silicon, the parallelization and distribution of the workload must be co-designed to minimize the impact of this degradation. Two load/efficiency masking examples (using collectives and prefetching) are mentioned above.
Third, with {\uGPU}s, the bandwidth and distance required from GPU-to-GPU links can be higher. Nevertheless, with optical links, we are looking towards petabit per second efficient communication across many racks, which is promising.

With regards to building an efficient, high-bandwidth {\uGPU} network, we have several options. First, as the traffic across {\uGPU}s that replace one large GPU is predictable, we can build a direct-connect topology within that group of {\uGPU}s and leave the remaining network as is. This is an approximation to the original network, though it eliminates the benefits of the smaller blast radius of {\uGPU}s. Alternatively, we can consider a (flat or hierarchical) switched network for the entire {\uGPU} cluster, yielding flexibility and improved fault-tolerance. Using circuit switching, in part or cluster-wide, may be crucial to achieve such a network at low cost. Circuit switching presents the following benefits over packet switching: (i) more than 50\% better energy efficiency, (ii) lower latency, and (iii) more ports at high bandwidth, which allows for larger and flatter networks~\cite{sirius}.

\paragraph{Data-center management} With {\uGPU}s, the number of devices per area is increased,
however, the energy per unit area is decreased.
There is active research to handle data-center management at scale using various automation techniques which can be applicable to {\uGPU} clusters~\cite{robots}.
Additionally, though the number of devices per rack may increase, the overall cooling requirements of the rack can be lighter due to the more efficient cooling of {\uGPU}s combined with co-packaged optics.
This can eliminate the need for liquid cooling racks in the data-center, which comprise a significant portion of racks, and thus space, in an NVIDIA B200 cluster~\cite{nvl72}.

\section{Case Study: LLM Inference}

\begin{figure*}[t!]
  \centering
  \begin{subfigure}[t]{0.96\columnwidth} 
    \centering
      \includegraphics[height=3.6cm]{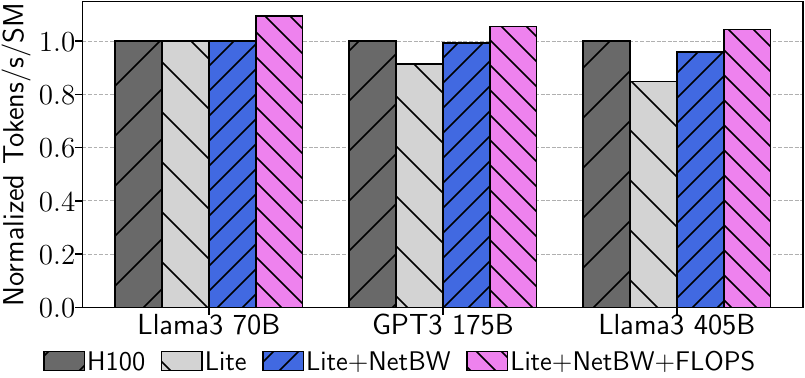}
      \caption{\textmd{Prompt prefill. All configurations perform similarly.
      As the model sizes grow, the ``{\uShort}'' cluster underperforms due to increased collectives causing network bottlenecks.
      Increasing the network bandwidth compensates the increased network demand, overclocking improves performance further as prefill workloads are compute-bound.}}
      \label{fig:prefill-perf}
  \end{subfigure}
  \hfill
  \begin{subfigure}[t]{1.05\columnwidth} 
    \centering
    \includegraphics[height=3.6cm]{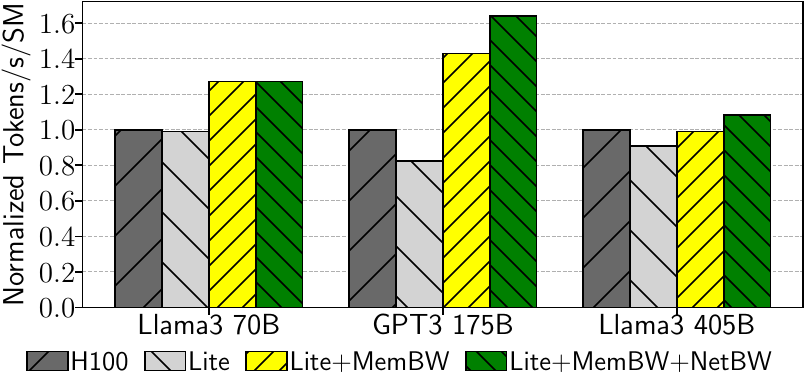}
    \caption{\textmd{Decode. As model sizes and thus the number of required GPUs grow, the ``{\uShort}'' cluster underperforms due to increased memory access intensities.
    The degradation is worse with GPT-3 due to it having more KV-heads resulting in proportionally longer memory-bound stages.
    As {\uGPU}s utilize their available shoreline for more memory bandwidth, performance improves and exceeds the current H100 cluster.}}
    \label{fig:decode-perf}
  \end{subfigure}
  \caption{\textmd{
    Results of the roofline modeling of H100 and {\uShort}-H100 clusters.
    Note that {\uShort}-H100 is already expected to be cheaper to manufacture, so we expect comparable performance to suffice.
  }}
  \label{fig:perf}
\end{figure*}

In this section, we present a case study of {\uGPU}s in the context of a trending AI workload –- LLM inference~\cite{attentionneed}.
LLM inference involves two distinct phases.
The prompt prefill phase processes input tokens to compute reusable intermediate states, i.e., the Key-Value (KV) cache, and generates the first new token.
The prefill phase is usually highly parallelizable and efficient in utilizing the compute resources.
The decode phase generates output tokens one at a time, with each new token building on the entire KV cache and appending to it.
This phase is often memory-bound and less efficient in compute utilization.
In the evaluation, we assume that different phases can execute on different {\uGPU} clusters~\cite{splitwise,distserve}
to demonstrate the hardware benefits achievable with {\uGPU}s.
With our case study on serving latest generative AI workloads, we aim to highlight potential advantages of {\uGPU}s that are modified from today's leading GPUs.

\paragraph{Methodology and workload}

We use roofline modeling~\cite{roofline} to capture important hardware and software characteristics and to model a {\uGPU} cluster running LLM inference.
We model important metrics including FLOPS, memory accesses, and the network traffic of collectives.
The modeling measures compute stages individually, including projection, MLP, and fused FlashAttention~\cite{fav3}.
Compute, memory I/O, and network I/O can overlap within each stage and tensor parallelism is used to distribute execution within each cluster.

NVIDIA H100 is the baseline GPU for comparison~\cite{H100specs}.
An H100 cluster consists of one to eight H100 GPUs. Each H100 includes 132 SMs.
The {\uGPU} is modeled based on H100 by reducing its capabilities to 1/4 of the original, denoted as ``{\uShort}'' in \Cref{tab:gpus}.
Accordingly, a {\uShort}-H100 cluster can consist of one to 32 {\uGPU}s, to match the total maximum number of SMs of the H100 cluster.
Recall that for {\uShort}-H100, we expect that bandwidth-to-compute can increase to $2\times$ of H100 and that
it can deliver higher sustainable FLOPS due to improved cooling efficiency.
To explore how these hardware improvements can impact performance,
we further define customized {\uGPU}s for comparison, as denoted and summarized in \Cref{tab:gpus}, with changed parameters highlighted in blue and red.

\setlength{\arrayrulewidth}{0.2pt}
\setlength{\tabcolsep}{2pt} 
\begin{table}[]
  \small
  \begin{center}
    \caption{\textmd{GPU configurations}}
    \label{tab:gpus}
    \vspace{-0.1in}
    \small
      \begin{tabular}{|c|c|c|c|c|c|}
      \hline
      GPU type & TFLOPS & Cap. & Mem BW & Net BW & \#Max \\
       &  & GB & GB/s & GB/s & GPUs \\
      \hline
      H100         & 2000        & 80         & 3352         & 450         & 8 \\
      \hdashline
      {\uShort}         & 500         & 20         & 838          & 112.5       & 32 \\
      \hdashline
      {\uShort}+NetBW       & 500         & 20         & 838          & \hspec{225} & 32 \\
      {\uShort}+NetBW+FLOPS & \hspec{550} & 20         & \lspec{419}  & \hspec{225} & 32 \\
      \hdashline
      {\uShort}+MemBW       & 500         & 20         & \hspec{1675} & 112.5       & 32 \\
      {\uShort}+MemBW+NetBW     & 500         & 20         & \hspec{1675} & \hspec{225} & 32 \\
      \hline
    \end{tabular}
  \end{center}
\end{table}

We evaluate performance with three LLM models with different sizes and structures: Llama3-70B, GPT3-175B, and Llama3-405B~\cite{llama,gpt3}.
We define the search criteria based on Splitwise's latency requirements, with TTFT (time-to-first-token) $\leq$ 1s and TBT (time-between-tokens) $\leq$ 50ms constraints~\cite{splitwise}.
We set a constant prompt sequence length of 1500 tokens, the reported median size in a production workload for coding~\cite{splitwise}.
The search sweeps all possible batch sizes and number of GPUs for each GPU type. Then, since different GPU types have different hardware capabilities, we normalize the throughput for each configuration using the number of SMs in that configuration. The resulting metric, throughput per SM (tokens/s/SM), represents the \emph{performance efficiency} of that configuration. For each GPU type, we plot the configuration with the highest throughput per SM.
Note that while we sweep up to the maximum number of GPUs per cluster as defined in \Cref{tab:gpus},
the search may return that running a model with less GPUs than the maximum yields better throughput per SM.

\paragraph{Results}
The results are summarized in \Cref{fig:perf}.
With this study, we show that while the basic {\uGPU} with no additional networking support could face performance limitations,
a {\uGPU} cluster can be customized to match or improve on the performance of a typical H100 cluster.
Note that this is in addition to the hardware and systems advantages of {\uGPU}s described in previous sections.
Additionally, note that customized and improved {\uGPU}s need not consume more energy at the cluster level,
as, e.g., they can trade-off FLOPS for bandwidth.

In terms of performance per \$-cost, which is the primary metric for cloud operators,
we expect the cost per comparable deployments to decrease with {\uGPU},
due to the improvements in the manufacturing cost of GPUs. In this case,
even matching performance of today's clusters may lead to sufficient improvement in performance per cost.
Nevertheless, the additional cost of networking needs consideration, and while it may be initially
a fraction of the GPU cost, it can turn into a bottleneck with increased scale.
Further analysis on performance and total cost of operation is vital for the viability of deploying {\uGPU}s at scale, though it is
out-of-scope for this paper.

\section{Related Work}

Running AI workloads on small chips has gained traction in the past years.
For example, Apple has been shipping Neural Engine in their mobile devices since 2017~\cite{neuralengine}.
Most recently, NVIDIA announced DIGITS as a powerful GPU workstation for engineering AI models prior to deployment on the cloud~\cite{digits}.
Also from the model design direction, improving inference for single GPUs has gained significant research attention~\cite{mllmnpu,infnpu,ondevice,powerinfer,phi}.
While these efforts aim to maximize AI capabilities on a single device, they do not address the challenges of scaling demanding AI workloads in the data-center.

On the other hand, Google's TPUs are an example of scaling AI workloads across many tensor processors~\cite{tpuv4}.
While they employ advanced networking technologies for lower cost and power consumption,
performance and flexibility limitations remain, such as a long reconfiguration periods and multi-device blast radii,
due to which a failure can render a group of TPUs inactive.
TPUs share similar principles with {\uGPU}s. However, TPUs are specialized and offer less programming flexibility compared to GPUs.
Additionally, TPUs have also packed more transistors into the same package across generations and are on a similar path to current complex GPUs~\cite{tpuv4, tpuv6e}.

Alternatively to the scale-out approach of {\uGPU}s, wafer-scale computing systems aim to pack massive amounts of compute and communication bandwidth onto single, large integrated chips~\cite{cerebras, wafer}. Though these systems benefit from much increased bandwidth and integration density, they require complex and advanced packaging techniques, which can lead to challenges around yield, cost, and power consumption~\cite{wafer}.

There is a plethora of work that propose systems solutions for improving the performance~\cite{sarathi,flashattention1}, energy efficiency~\cite{userve,greener},
parallelism~\cite{megatron,ringattn}, and scheduling~\cite{sched,llumnix} of AI workloads in the data-center. Recently, DeepSeek demonstrated a variety of optimizations that enable efficient training and serving of a strong LLM on hardware that are relatively weaker than cutting-edge GPUs~\cite{deepseekv3}. These works are complementary to the hardware and systems efforts on delivering cost-effective scaling of AI workloads using {\uGPU}s.

\section{Conclusion}
We are already facing uncertainty on the amount of compute and memory that can fit
into a single GPU package, as cutting-edge GPUs already display the 
packaging, cooling, power- and cost-related challenges of their complex designs.
In this paper, we propose an alternative way of scaling AI infrastructure: 
by using {\uGPU}s instead of complex and expensive large GPUs. 
Motivated by the yield, power, and operational benefits of 
smaller GPU packages, we look at AI infrastructure within the context of {\uGPU}s.
We provide an overview of key research questions around workload, memory, and network management.
We also present how {\uGPU}s can improve energy management, performance efficiency, and fault-tolerance.
With this paper, we
aim to start a discussion around {\uGPU}s and their potential to turn the tide on the many issues
we face while building and operating GPU clusters in the era of generative AI.



\bibliographystyle{ACM-Reference-Format}
\bibliography{refs}


\begin{thebibliography}{63}


\ifx \showCODEN    \undefined \def \showCODEN     #1{\unskip}     \fi
\ifx \showISBNx    \undefined \def \showISBNx     #1{\unskip}     \fi
\ifx \showISBNxiii \undefined \def \showISBNxiii  #1{\unskip}     \fi
\ifx \showISSN     \undefined \def \showISSN      #1{\unskip}     \fi
\ifx \showLCCN     \undefined \def \showLCCN      #1{\unskip}     \fi
\ifx \shownote     \undefined \def \shownote      #1{#1}          \fi
\ifx \showarticletitle \undefined \def \showarticletitle #1{#1}   \fi
\ifx \showURL      \undefined \def \showURL       {\relax}        \fi
\providecommand\bibfield[2]{#2}
\providecommand\bibinfo[2]{#2}
\providecommand\natexlab[1]{#1}
\providecommand\showeprint[2][]{arXiv:#2}

\bibitem[nvl({[n.\,d.]})]%
        {nvl72}
 \bibinfo{year}{[n.\,d.]}\natexlab{}.
\newblock \bibinfo{title}{NVIDIA GB200 NVL72}.
\newblock
  \bibinfo{howpublished}{\url{https://www.nvidia.com/en-us/data-center/gb200-nvl72/}}.
\newblock
\newblock
\shownote{Accessed: 2025-01-15}.


\bibitem[ope(2024)]%
        {openai_infra}
 \bibinfo{year}{2024}\natexlab{}.
\newblock \showarticletitle{Hot Chips 2024 Conference Proceedings}.
\newblock
\urldef\tempurl%
\url{https://hc2024.hotchips.org/assets/program/conference/day1/HotChips%20-%202024-08-26.pdf}
\showURL{%
\tempurl}
\newblock
\shownote{Accessed: 2025-01-16}.


\bibitem[Abdin et~al\mbox{.}(2024)]%
        {phi}
\bibfield{author}{\bibinfo{person}{Marah Abdin}, \bibinfo{person}{Jyoti Aneja},
  \bibinfo{person}{Hany Awadalla}, \bibinfo{person}{Ahmed Awadallah},
  \bibinfo{person}{Ammar~Ahmad Awan}, \bibinfo{person}{Nguyen Bach},
  \bibinfo{person}{Amit Bahree}, \bibinfo{person}{Arash Bakhtiari},
  \bibinfo{person}{Jianmin Bao}, \bibinfo{person}{Harkirat Behl},
  \bibinfo{person}{Alon Benhaim}, \bibinfo{person}{Misha Bilenko},
  \bibinfo{person}{Johan Bjorck}, \bibinfo{person}{Sébastien Bubeck},
  \bibinfo{person}{Martin Cai}, \bibinfo{person}{Qin Cai},
  \bibinfo{person}{Vishrav Chaudhary}, \bibinfo{person}{Dong Chen},
  \bibinfo{person}{Dongdong Chen}, \bibinfo{person}{Weizhu Chen},
  \bibinfo{person}{Yen-Chun Chen}, \bibinfo{person}{Yi-Ling Chen},
  \bibinfo{person}{Hao Cheng}, \bibinfo{person}{Parul Chopra},
  \bibinfo{person}{Xiyang Dai}, \bibinfo{person}{Matthew Dixon},
  \bibinfo{person}{Ronen Eldan}, \bibinfo{person}{Victor Fragoso},
  \bibinfo{person}{Jianfeng Gao}, \bibinfo{person}{Mei Gao},
  \bibinfo{person}{Min Gao}, \bibinfo{person}{Amit Garg},
  \bibinfo{person}{Allie~Del Giorno}, \bibinfo{person}{Abhishek Goswami},
  \bibinfo{person}{Suriya Gunasekar}, \bibinfo{person}{Emman Haider},
  \bibinfo{person}{Junheng Hao}, \bibinfo{person}{Russell~J. Hewett},
  \bibinfo{person}{Wenxiang Hu}, \bibinfo{person}{Jamie Huynh},
  \bibinfo{person}{Dan Iter}, \bibinfo{person}{Sam~Ade Jacobs},
  \bibinfo{person}{Mojan Javaheripi}, \bibinfo{person}{Xin Jin},
  \bibinfo{person}{Nikos Karampatziakis}, \bibinfo{person}{Piero Kauffmann},
  \bibinfo{person}{Mahoud Khademi}, \bibinfo{person}{Dongwoo Kim},
  \bibinfo{person}{Young~Jin Kim}, \bibinfo{person}{Lev Kurilenko},
  \bibinfo{person}{James~R. Lee}, \bibinfo{person}{Yin~Tat Lee},
  \bibinfo{person}{Yuanzhi Li}, \bibinfo{person}{Yunsheng Li},
  \bibinfo{person}{Chen Liang}, \bibinfo{person}{Lars Liden},
  \bibinfo{person}{Xihui Lin}, \bibinfo{person}{Zeqi Lin}, \bibinfo{person}{Ce
  Liu}, \bibinfo{person}{Liyuan Liu}, \bibinfo{person}{Mengchen Liu},
  \bibinfo{person}{Weishung Liu}, \bibinfo{person}{Xiaodong Liu},
  \bibinfo{person}{Chong Luo}, \bibinfo{person}{Piyush Madan},
  \bibinfo{person}{Ali Mahmoudzadeh}, \bibinfo{person}{David Majercak},
  \bibinfo{person}{Matt Mazzola}, \bibinfo{person}{Caio César~Teodoro Mendes},
  \bibinfo{person}{Arindam Mitra}, \bibinfo{person}{Hardik Modi},
  \bibinfo{person}{Anh Nguyen}, \bibinfo{person}{Brandon Norick},
  \bibinfo{person}{Barun Patra}, \bibinfo{person}{Daniel Perez-Becker},
  \bibinfo{person}{Thomas Portet}, \bibinfo{person}{Reid Pryzant},
  \bibinfo{person}{Heyang Qin}, \bibinfo{person}{Marko Radmilac},
  \bibinfo{person}{Liliang Ren}, \bibinfo{person}{Gustavo de Rosa},
  \bibinfo{person}{Corby Rosset}, \bibinfo{person}{Sambudha Roy},
  \bibinfo{person}{Olatunji Ruwase}, \bibinfo{person}{Olli Saarikivi},
  \bibinfo{person}{Amin Saied}, \bibinfo{person}{Adil Salim},
  \bibinfo{person}{Michael Santacroce}, \bibinfo{person}{Shital Shah},
  \bibinfo{person}{Ning Shang}, \bibinfo{person}{Hiteshi Sharma},
  \bibinfo{person}{Yelong Shen}, \bibinfo{person}{Swadheen Shukla},
  \bibinfo{person}{Xia Song}, \bibinfo{person}{Masahiro Tanaka},
  \bibinfo{person}{Andrea Tupini}, \bibinfo{person}{Praneetha Vaddamanu},
  \bibinfo{person}{Chunyu Wang}, \bibinfo{person}{Guanhua Wang},
  \bibinfo{person}{Lijuan Wang}, \bibinfo{person}{Shuohang Wang},
  \bibinfo{person}{Xin Wang}, \bibinfo{person}{Yu Wang},
  \bibinfo{person}{Rachel Ward}, \bibinfo{person}{Wen Wen},
  \bibinfo{person}{Philipp Witte}, \bibinfo{person}{Haiping Wu},
  \bibinfo{person}{Xiaoxia Wu}, \bibinfo{person}{Michael Wyatt},
  \bibinfo{person}{Bin Xiao}, \bibinfo{person}{Can Xu},
  \bibinfo{person}{Jiahang Xu}, \bibinfo{person}{Weijian Xu},
  \bibinfo{person}{Jilong Xue}, \bibinfo{person}{Sonali Yadav},
  \bibinfo{person}{Fan Yang}, \bibinfo{person}{Jianwei Yang},
  \bibinfo{person}{Yifan Yang}, \bibinfo{person}{Ziyi Yang},
  \bibinfo{person}{Donghan Yu}, \bibinfo{person}{Lu Yuan},
  \bibinfo{person}{Chenruidong Zhang}, \bibinfo{person}{Cyril Zhang},
  \bibinfo{person}{Jianwen Zhang}, \bibinfo{person}{Li~Lyna Zhang},
  \bibinfo{person}{Yi Zhang}, \bibinfo{person}{Yue Zhang},
  \bibinfo{person}{Yunan Zhang}, {and} \bibinfo{person}{Xiren Zhou}.}
  \bibinfo{year}{2024}\natexlab{}.
\newblock \bibinfo{title}{Phi-3 Technical Report: A Highly Capable Language
  Model Locally on Your Phone}.
\newblock
\showeprint[arxiv]{2404.14219}~[cs.CL]
\urldef\tempurl%
\url{https://arxiv.org/abs/2404.14219}
\showURL{%
\tempurl}


\bibitem[Agrawal et~al\mbox{.}(2023)]%
        {sarathi}
\bibfield{author}{\bibinfo{person}{Amey Agrawal}, \bibinfo{person}{Ashish
  Panwar}, \bibinfo{person}{Jayashree Mohan}, \bibinfo{person}{Nipun Kwatra},
  \bibinfo{person}{Bhargav~S. Gulavani}, {and} \bibinfo{person}{Ramachandran
  Ramjee}.} \bibinfo{year}{2023}\natexlab{}.
\newblock \bibinfo{title}{SARATHI: Efficient LLM Inference by Piggybacking
  Decodes with Chunked Prefills}.
\newblock
\showeprint[arxiv]{2308.16369}~[cs.LG]
\urldef\tempurl%
\url{https://arxiv.org/abs/2308.16369}
\showURL{%
\tempurl}


\bibitem[Aminabadi et~al\mbox{.}(2022)]%
        {deepinf}
\bibfield{author}{\bibinfo{person}{Reza~Yazdani Aminabadi},
  \bibinfo{person}{Samyam Rajbhandari}, \bibinfo{person}{Minjia Zhang},
  \bibinfo{person}{Ammar~Ahmad Awan}, \bibinfo{person}{Cheng Li},
  \bibinfo{person}{Du Li}, \bibinfo{person}{Elton Zheng}, \bibinfo{person}{Jeff
  Rasley}, \bibinfo{person}{Shaden Smith}, \bibinfo{person}{Olatunji Ruwase},
  {and} \bibinfo{person}{Yuxiong He}.} \bibinfo{year}{2022}\natexlab{}.
\newblock \bibinfo{title}{DeepSpeed Inference: Enabling Efficient Inference of
  Transformer Models at Unprecedented Scale}.
\newblock
\showeprint[arxiv]{2207.00032}~[cs.LG]
\urldef\tempurl%
\url{https://arxiv.org/abs/2207.00032}
\showURL{%
\tempurl}


\bibitem[Ballani et~al\mbox{.}(2020)]%
        {sirius}
\bibfield{author}{\bibinfo{person}{Hitesh Ballani}, \bibinfo{person}{Paolo
  Costa}, \bibinfo{person}{Raphael Behrendt}, \bibinfo{person}{Daniel
  Cletheroe}, \bibinfo{person}{Istvan Haller}, \bibinfo{person}{Krzysztof
  Jozwik}, \bibinfo{person}{Fotini Karinou}, \bibinfo{person}{Sophie Lange},
  \bibinfo{person}{Kai Shi}, \bibinfo{person}{Benn Thomsen}, {and}
  \bibinfo{person}{Hugh Williams}.} \bibinfo{year}{2020}\natexlab{}.
\newblock \showarticletitle{Sirius: A Flat Datacenter Network with Nanosecond
  Optical Switching}. In \bibinfo{booktitle}{\emph{Proceedings of the Annual
  Conference of the ACM Special Interest Group on Data Communication on the
  Applications, Technologies, Architectures, and Protocols for Computer
  Communication}} (Virtual Event, USA) \emph{(\bibinfo{series}{SIGCOMM '20})}.
  \bibinfo{publisher}{Association for Computing Machinery},
  \bibinfo{address}{New York, NY, USA}, \bibinfo{pages}{782–797}.
\newblock
\showISBNx{9781450379557}
\href{https://doi.org/10.1145/3387514.3406221}{doi:\nolinkurl{10.1145/3387514.3406221}}


\bibitem[Brown et~al\mbox{.}(2020)]%
        {gpt3}
\bibfield{author}{\bibinfo{person}{Tom~B. Brown}, \bibinfo{person}{Benjamin
  Mann}, \bibinfo{person}{Nick Ryder}, \bibinfo{person}{Melanie Subbiah},
  \bibinfo{person}{Jared Kaplan}, \bibinfo{person}{Prafulla Dhariwal},
  \bibinfo{person}{Arvind Neelakantan}, \bibinfo{person}{Pranav Shyam},
  \bibinfo{person}{Girish Sastry}, \bibinfo{person}{Amanda Askell},
  \bibinfo{person}{Sandhini Agarwal}, \bibinfo{person}{Ariel Herbert-Voss},
  \bibinfo{person}{Gretchen Krueger}, \bibinfo{person}{Tom Henighan},
  \bibinfo{person}{Rewon Child}, \bibinfo{person}{Aditya Ramesh},
  \bibinfo{person}{Daniel~M. Ziegler}, \bibinfo{person}{Jeffrey Wu},
  \bibinfo{person}{Clemens Winter}, \bibinfo{person}{Christopher Hesse},
  \bibinfo{person}{Mark Chen}, \bibinfo{person}{Eric Sigler},
  \bibinfo{person}{Mateusz Litwin}, \bibinfo{person}{Scott Gray},
  \bibinfo{person}{Benjamin Chess}, \bibinfo{person}{Jack Clark},
  \bibinfo{person}{Christopher Berner}, \bibinfo{person}{Sam McCandlish},
  \bibinfo{person}{Alec Radford}, \bibinfo{person}{Ilya Sutskever}, {and}
  \bibinfo{person}{Dario Amodei}.} \bibinfo{year}{2020}\natexlab{}.
\newblock \bibinfo{title}{Language Models are Few-Shot Learners}.
\newblock
\showeprint[arxiv]{2005.14165}~[cs.CL]
\urldef\tempurl%
\url{https://arxiv.org/abs/2005.14165}
\showURL{%
\tempurl}


\bibitem[Cerebras({[n.\,d.]})]%
        {cerebras}
\bibfield{author}{\bibinfo{person}{Cerebras}.}
  \bibinfo{year}{[n.\,d.]}\natexlab{}.
\newblock \bibinfo{title}{Wafer Scale Engine 3}.
\newblock \bibinfo{howpublished}{\url{https://www.cerebras.ai/chip}}.
\newblock
\newblock
\shownote{[Accessed 17-04-2025]}.


\bibitem[Chung et~al\mbox{.}(2024)]%
        {perseus}
\bibfield{author}{\bibinfo{person}{Jae-Won Chung}, \bibinfo{person}{Yile Gu},
  \bibinfo{person}{Insu Jang}, \bibinfo{person}{Luoxi Meng},
  \bibinfo{person}{Nikhil Bansal}, {and} \bibinfo{person}{Mosharaf Chowdhury}.}
  \bibinfo{year}{2024}\natexlab{}.
\newblock \showarticletitle{Reducing Energy Bloat in Large Model Training}. In
  \bibinfo{booktitle}{\emph{Proceedings of the ACM SIGOPS 30th Symposium on
  Operating Systems Principles}} (Austin, TX, USA) \emph{(\bibinfo{series}{SOSP
  '24})}. \bibinfo{publisher}{Association for Computing Machinery},
  \bibinfo{address}{New York, NY, USA}, \bibinfo{pages}{144–159}.
\newblock
\showISBNx{9798400712517}
\href{https://doi.org/10.1145/3694715.3695970}{doi:\nolinkurl{10.1145/3694715.3695970}}


\bibitem[Cloud(2025)]%
        {tpuv6e}
\bibfield{author}{\bibinfo{person}{Google Cloud}.}
  \bibinfo{year}{2025}\natexlab{}.
\newblock \bibinfo{title}{TPU v6e}.
\newblock
\urldef\tempurl%
\url{https://cloud.google.com/tpu/docs/v6e}
\showURL{%
\tempurl}
\newblock
\shownote{Accessed: 2025-01-15}.


\bibitem[Corporation(2022)]%
        {H100specs}
\bibfield{author}{\bibinfo{person}{NVIDIA Corporation}.}
  \bibinfo{year}{2022}\natexlab{}.
\newblock \bibinfo{title}{NVIDIA H100 Tensor Core GPU Architecture Overview}.
\newblock
\urldef\tempurl%
\url{https://resources.nvidia.com/en-us-tensor-core/gtc22-whitepaper-hopper}
\showURL{%
\tempurl}
\newblock
\shownote{Accessed: 2025-01-10}.


\bibitem[Corporation(2025)]%
        {thottle_heat2}
\bibfield{author}{\bibinfo{person}{NVIDIA Corporation}.}
  \bibinfo{year}{2025}\natexlab{}.
\newblock \bibinfo{title}{NVIDIA H100 NVL GPU}.
\newblock
\urldef\tempurl%
\url{https://www.nvidia.com/content/dam/en-zz/Solutions/Data-Center/h100/PB-11773-001_v01.pdf}
\showURL{%
\tempurl}
\newblock
\shownote{Accessed: 2025-01-15}.


\bibitem[Dao et~al\mbox{.}(2022)]%
        {flashattention1}
\bibfield{author}{\bibinfo{person}{Tri Dao}, \bibinfo{person}{Dan Fu},
  \bibinfo{person}{Stefano Ermon}, \bibinfo{person}{Atri Rudra}, {and}
  \bibinfo{person}{Christopher R{\'e}}.} \bibinfo{year}{2022}\natexlab{}.
\newblock \showarticletitle{Flashattention: Fast and memory-efficient exact
  attention with io-awareness}.
\newblock \bibinfo{journal}{\emph{Advances in Neural Information Processing
  Systems}}  \bibinfo{volume}{35} (\bibinfo{year}{2022}),
  \bibinfo{pages}{16344--16359}.
\newblock


\bibitem[DeepSeek-AI et~al\mbox{.}(2025)]%
        {deepseekv3}
\bibfield{author}{\bibinfo{person}{DeepSeek-AI}, \bibinfo{person}{Aixin Liu},
  \bibinfo{person}{Bei Feng}, \bibinfo{person}{Bing Xue},
  \bibinfo{person}{Bingxuan Wang}, \bibinfo{person}{Bochao Wu},
  \bibinfo{person}{Chengda Lu}, \bibinfo{person}{Chenggang Zhao},
  \bibinfo{person}{Chengqi Deng}, \bibinfo{person}{Chenyu Zhang},
  \bibinfo{person}{Chong Ruan}, \bibinfo{person}{Damai Dai},
  \bibinfo{person}{Daya Guo}, \bibinfo{person}{Dejian Yang},
  \bibinfo{person}{Deli Chen}, \bibinfo{person}{Dongjie Ji},
  \bibinfo{person}{Erhang Li}, \bibinfo{person}{Fangyun Lin},
  \bibinfo{person}{Fucong Dai}, \bibinfo{person}{Fuli Luo},
  \bibinfo{person}{Guangbo Hao}, \bibinfo{person}{Guanting Chen},
  \bibinfo{person}{Guowei Li}, \bibinfo{person}{H. Zhang}, \bibinfo{person}{Han
  Bao}, \bibinfo{person}{Hanwei Xu}, \bibinfo{person}{Haocheng Wang},
  \bibinfo{person}{Haowei Zhang}, \bibinfo{person}{Honghui Ding},
  \bibinfo{person}{Huajian Xin}, \bibinfo{person}{Huazuo Gao},
  \bibinfo{person}{Hui Li}, \bibinfo{person}{Hui Qu}, \bibinfo{person}{J.~L.
  Cai}, \bibinfo{person}{Jian Liang}, \bibinfo{person}{Jianzhong Guo},
  \bibinfo{person}{Jiaqi Ni}, \bibinfo{person}{Jiashi Li},
  \bibinfo{person}{Jiawei Wang}, \bibinfo{person}{Jin Chen},
  \bibinfo{person}{Jingchang Chen}, \bibinfo{person}{Jingyang Yuan},
  \bibinfo{person}{Junjie Qiu}, \bibinfo{person}{Junlong Li},
  \bibinfo{person}{Junxiao Song}, \bibinfo{person}{Kai Dong},
  \bibinfo{person}{Kai Hu}, \bibinfo{person}{Kaige Gao}, \bibinfo{person}{Kang
  Guan}, \bibinfo{person}{Kexin Huang}, \bibinfo{person}{Kuai Yu},
  \bibinfo{person}{Lean Wang}, \bibinfo{person}{Lecong Zhang},
  \bibinfo{person}{Lei Xu}, \bibinfo{person}{Leyi Xia}, \bibinfo{person}{Liang
  Zhao}, \bibinfo{person}{Litong Wang}, \bibinfo{person}{Liyue Zhang},
  \bibinfo{person}{Meng Li}, \bibinfo{person}{Miaojun Wang},
  \bibinfo{person}{Mingchuan Zhang}, \bibinfo{person}{Minghua Zhang},
  \bibinfo{person}{Minghui Tang}, \bibinfo{person}{Mingming Li},
  \bibinfo{person}{Ning Tian}, \bibinfo{person}{Panpan Huang},
  \bibinfo{person}{Peiyi Wang}, \bibinfo{person}{Peng Zhang},
  \bibinfo{person}{Qiancheng Wang}, \bibinfo{person}{Qihao Zhu},
  \bibinfo{person}{Qinyu Chen}, \bibinfo{person}{Qiushi Du},
  \bibinfo{person}{R.~J. Chen}, \bibinfo{person}{R.~L. Jin},
  \bibinfo{person}{Ruiqi Ge}, \bibinfo{person}{Ruisong Zhang},
  \bibinfo{person}{Ruizhe Pan}, \bibinfo{person}{Runji Wang},
  \bibinfo{person}{Runxin Xu}, \bibinfo{person}{Ruoyu Zhang},
  \bibinfo{person}{Ruyi Chen}, \bibinfo{person}{S.~S. Li},
  \bibinfo{person}{Shanghao Lu}, \bibinfo{person}{Shangyan Zhou},
  \bibinfo{person}{Shanhuang Chen}, \bibinfo{person}{Shaoqing Wu},
  \bibinfo{person}{Shengfeng Ye}, \bibinfo{person}{Shengfeng Ye},
  \bibinfo{person}{Shirong Ma}, \bibinfo{person}{Shiyu Wang},
  \bibinfo{person}{Shuang Zhou}, \bibinfo{person}{Shuiping Yu},
  \bibinfo{person}{Shunfeng Zhou}, \bibinfo{person}{Shuting Pan},
  \bibinfo{person}{T. Wang}, \bibinfo{person}{Tao Yun}, \bibinfo{person}{Tian
  Pei}, \bibinfo{person}{Tianyu Sun}, \bibinfo{person}{W.~L. Xiao},
  \bibinfo{person}{Wangding Zeng}, \bibinfo{person}{Wanjia Zhao},
  \bibinfo{person}{Wei An}, \bibinfo{person}{Wen Liu}, \bibinfo{person}{Wenfeng
  Liang}, \bibinfo{person}{Wenjun Gao}, \bibinfo{person}{Wenqin Yu},
  \bibinfo{person}{Wentao Zhang}, \bibinfo{person}{X.~Q. Li},
  \bibinfo{person}{Xiangyue Jin}, \bibinfo{person}{Xianzu Wang},
  \bibinfo{person}{Xiao Bi}, \bibinfo{person}{Xiaodong Liu},
  \bibinfo{person}{Xiaohan Wang}, \bibinfo{person}{Xiaojin Shen},
  \bibinfo{person}{Xiaokang Chen}, \bibinfo{person}{Xiaokang Zhang},
  \bibinfo{person}{Xiaosha Chen}, \bibinfo{person}{Xiaotao Nie},
  \bibinfo{person}{Xiaowen Sun}, \bibinfo{person}{Xiaoxiang Wang},
  \bibinfo{person}{Xin Cheng}, \bibinfo{person}{Xin Liu}, \bibinfo{person}{Xin
  Xie}, \bibinfo{person}{Xingchao Liu}, \bibinfo{person}{Xingkai Yu},
  \bibinfo{person}{Xinnan Song}, \bibinfo{person}{Xinxia Shan},
  \bibinfo{person}{Xinyi Zhou}, \bibinfo{person}{Xinyu Yang},
  \bibinfo{person}{Xinyuan Li}, \bibinfo{person}{Xuecheng Su},
  \bibinfo{person}{Xuheng Lin}, \bibinfo{person}{Y.~K. Li},
  \bibinfo{person}{Y.~Q. Wang}, \bibinfo{person}{Y.~X. Wei},
  \bibinfo{person}{Y.~X. Zhu}, \bibinfo{person}{Yang Zhang},
  \bibinfo{person}{Yanhong Xu}, \bibinfo{person}{Yanhong Xu},
  \bibinfo{person}{Yanping Huang}, \bibinfo{person}{Yao Li},
  \bibinfo{person}{Yao Zhao}, \bibinfo{person}{Yaofeng Sun},
  \bibinfo{person}{Yaohui Li}, \bibinfo{person}{Yaohui Wang},
  \bibinfo{person}{Yi Yu}, \bibinfo{person}{Yi Zheng}, \bibinfo{person}{Yichao
  Zhang}, \bibinfo{person}{Yifan Shi}, \bibinfo{person}{Yiliang Xiong},
  \bibinfo{person}{Ying He}, \bibinfo{person}{Ying Tang},
  \bibinfo{person}{Yishi Piao}, \bibinfo{person}{Yisong Wang},
  \bibinfo{person}{Yixuan Tan}, \bibinfo{person}{Yiyang Ma},
  \bibinfo{person}{Yiyuan Liu}, \bibinfo{person}{Yongqiang Guo},
  \bibinfo{person}{Yu Wu}, \bibinfo{person}{Yuan Ou}, \bibinfo{person}{Yuchen
  Zhu}, \bibinfo{person}{Yuduan Wang}, \bibinfo{person}{Yue Gong},
  \bibinfo{person}{Yuheng Zou}, \bibinfo{person}{Yujia He},
  \bibinfo{person}{Yukun Zha}, \bibinfo{person}{Yunfan Xiong},
  \bibinfo{person}{Yunxian Ma}, \bibinfo{person}{Yuting Yan},
  \bibinfo{person}{Yuxiang Luo}, \bibinfo{person}{Yuxiang You},
  \bibinfo{person}{Yuxuan Liu}, \bibinfo{person}{Yuyang Zhou},
  \bibinfo{person}{Z.~F. Wu}, \bibinfo{person}{Z.~Z. Ren},
  \bibinfo{person}{Zehui Ren}, \bibinfo{person}{Zhangli Sha},
  \bibinfo{person}{Zhe Fu}, \bibinfo{person}{Zhean Xu}, \bibinfo{person}{Zhen
  Huang}, \bibinfo{person}{Zhen Zhang}, \bibinfo{person}{Zhenda Xie},
  \bibinfo{person}{Zhengyan Zhang}, \bibinfo{person}{Zhewen Hao},
  \bibinfo{person}{Zhibin Gou}, \bibinfo{person}{Zhicheng Ma},
  \bibinfo{person}{Zhigang Yan}, \bibinfo{person}{Zhihong Shao},
  \bibinfo{person}{Zhipeng Xu}, \bibinfo{person}{Zhiyu Wu},
  \bibinfo{person}{Zhongyu Zhang}, \bibinfo{person}{Zhuoshu Li},
  \bibinfo{person}{Zihui Gu}, \bibinfo{person}{Zijia Zhu},
  \bibinfo{person}{Zijun Liu}, \bibinfo{person}{Zilin Li},
  \bibinfo{person}{Ziwei Xie}, \bibinfo{person}{Ziyang Song},
  \bibinfo{person}{Ziyi Gao}, {and} \bibinfo{person}{Zizheng Pan}.}
  \bibinfo{year}{2025}\natexlab{}.
\newblock \bibinfo{title}{DeepSeek-V3 Technical Report}.
\newblock
\showeprint[arxiv]{2412.19437}~[cs.CL]
\urldef\tempurl%
\url{https://arxiv.org/abs/2412.19437}
\showURL{%
\tempurl}


\bibitem[der Wijngaart and Oh(2022)]%
        {gpuprefetch}
\bibfield{author}{\bibinfo{person}{Rob~Van der Wijngaart} {and}
  \bibinfo{person}{Fred Oh}.} \bibinfo{year}{2022}\natexlab{}.
\newblock \bibinfo{title}{Boosting Application Performance with GPU Memory
  Prefetching}.
\newblock
\urldef\tempurl%
\url{https://developer.nvidia.com/blog/boosting-application-performance-with-gpu-memory-prefetching/}
\showURL{%
\tempurl}
\newblock
\shownote{Accessed: 2025-01-15}.


\bibitem[Face({[n.\,d.]})]%
        {llama_infer}
\bibfield{author}{\bibinfo{person}{Hugging Face}.}
  \bibinfo{year}{[n.\,d.]}\natexlab{}.
\newblock \bibinfo{title}{{L}lama 3.1 - 405{B}, 70{B} \& 8{B} with
  multilinguality and long context}.
\newblock \bibinfo{howpublished}{\url{https://huggingface.co/blog/llama31}}.
\newblock
\newblock
\shownote{[Accessed 15-04-2025]}.


\bibitem[Fu et~al\mbox{.}(2024)]%
        {sched}
\bibfield{author}{\bibinfo{person}{Yichao Fu}, \bibinfo{person}{Siqi Zhu},
  \bibinfo{person}{Runlong Su}, \bibinfo{person}{Aurick Qiao},
  \bibinfo{person}{Ion Stoica}, {and} \bibinfo{person}{Hao Zhang}.}
  \bibinfo{year}{2024}\natexlab{}.
\newblock \bibinfo{title}{Efficient LLM Scheduling by Learning to Rank}.
\newblock
\showeprint[arxiv]{2408.15792}~[cs.LG]
\urldef\tempurl%
\url{https://arxiv.org/abs/2408.15792}
\showURL{%
\tempurl}


\bibitem[Garrou(2024)]%
        {cowos}
\bibfield{author}{\bibinfo{person}{Phil Garrou}.}
  \bibinfo{year}{2024}\natexlab{}.
\newblock \showarticletitle{IFTLE 607: Why Nvidia’s Blackwell is Having
  Issues with TSMC CoWoS-L Technology}.
\newblock \bibinfo{journal}{\emph{3DInCites}} (\bibinfo{year}{2024}).
\newblock
\urldef\tempurl%
\url{{https://www.3dincites.com/2024/10/iftle-607-why-nvidias-blackwell-is-having-issues-with-tsmc-cowos-l-technology/}}
\showURL{%
\tempurl}
\newblock
\shownote{Accessed: 2025-01-14}.


\bibitem[Gupta and Lathrop(1972)]%
        {yield1}
\bibfield{author}{\bibinfo{person}{A. Gupta} {and} \bibinfo{person}{J.W.
  Lathrop}.} \bibinfo{year}{1972}\natexlab{}.
\newblock \showarticletitle{Yield analysis of large integrated-circuit chips}.
\newblock \bibinfo{journal}{\emph{IEEE Journal of Solid-State Circuits}}
  \bibinfo{volume}{7}, \bibinfo{number}{5} (\bibinfo{year}{1972}),
  \bibinfo{pages}{389--395}.
\newblock
\href{https://doi.org/10.1109/JSSC.1972.1052898}{doi:\nolinkurl{10.1109/JSSC.1972.1052898}}


\bibitem[He(2024)]%
        {thottle_heat}
\bibfield{author}{\bibinfo{person}{Horace He}.}
  \bibinfo{year}{2024}\natexlab{}.
\newblock \bibinfo{title}{Strangely, Matrix Multiplications on GPUs Run Faster
  When Given "Predictable" Data!}
\newblock
\urldef\tempurl%
\url{https://www.thonking.ai/p/strangely-matrix-multiplications}
\showURL{%
\tempurl}
\newblock
\shownote{Accessed: 2025-01-15}.


\bibitem[Heydari et~al\mbox{.}(2022)]%
        {liquid1}
\bibfield{author}{\bibinfo{person}{Ali Heydari}, \bibinfo{person}{Pardeep
  Shahi}, \bibinfo{person}{Vahideh Radmard}, \bibinfo{person}{Bahareh Eslami},
  \bibinfo{person}{Uschas Chowdhury}, \bibinfo{person}{Satyam Saini},
  \bibinfo{person}{Pratik Bansode}, \bibinfo{person}{Harold Miyamura},
  \bibinfo{person}{Dereje Agonafer}, {and} \bibinfo{person}{Jeremy Rodriguez}.}
  \bibinfo{year}{2022}\natexlab{}.
\newblock \showarticletitle{Liquid to Liquid Cooling for High Heat Density
  Liquid Cooled Data Centers}. In \bibinfo{booktitle}{\emph{International
  Electronic Packaging Technical Conference and Exhibition}},
  Vol.~\bibinfo{volume}{86557}. American Society of Mechanical Engineers,
  \bibinfo{pages}{V001T01A007}.
\newblock


\bibitem[Hong et~al\mbox{.}(2024)]%
        {robots}
\bibfield{author}{\bibinfo{person}{Freddie Hong}, \bibinfo{person}{Iason
  Sarantopoulos}, \bibinfo{person}{Elliott Hogg}, \bibinfo{person}{David
  Richardson}, \bibinfo{person}{Yizhong Zhang}, \bibinfo{person}{Hugh
  Williams}, \bibinfo{person}{David Sweeney}, \bibinfo{person}{Andromachi
  Chatzieleftheriou}, {and} \bibinfo{person}{Antony Rowstron}.}
  \bibinfo{year}{2024}\natexlab{}.
\newblock \showarticletitle{Self-maintaining [networked] systems: The rise of
  datacenter robotics!}. In \bibinfo{booktitle}{\emph{Proceedings of the 23rd
  ACM Workshop on Hot Topics in Networks}} (Irvine, CA, USA)
  \emph{(\bibinfo{series}{HotNets '24})}. \bibinfo{publisher}{Association for
  Computing Machinery}, \bibinfo{address}{New York, NY, USA},
  \bibinfo{pages}{159–166}.
\newblock
\showISBNx{9798400712722}
\href{https://doi.org/10.1145/3696348.3696872}{doi:\nolinkurl{10.1145/3696348.3696872}}


\bibitem[Hu et~al\mbox{.}(2024)]%
        {wafer}
\bibfield{author}{\bibinfo{person}{Yang Hu}, \bibinfo{person}{Xinhan Lin},
  \bibinfo{person}{Huizheng Wang}, \bibinfo{person}{Zhen He},
  \bibinfo{person}{Xingmao Yu}, \bibinfo{person}{Jiahao Zhang},
  \bibinfo{person}{Qize Yang}, \bibinfo{person}{Zheng Xu},
  \bibinfo{person}{Sihan Guan}, \bibinfo{person}{Jiahao Fang},
  \bibinfo{person}{Haoran Shang}, \bibinfo{person}{Xinru Tang},
  \bibinfo{person}{Xu Dai}, \bibinfo{person}{Shaojun Wei}, {and}
  \bibinfo{person}{Shouyi Yin}.} \bibinfo{year}{2024}\natexlab{}.
\newblock \showarticletitle{Wafer-Scale Computing: Advancements, Challenges,
  and Future Perspectives [Feature]}.
\newblock \bibinfo{journal}{\emph{IEEE Circuits and Systems Magazine}}
  \bibinfo{volume}{24}, \bibinfo{number}{1} (\bibinfo{year}{2024}),
  \bibinfo{pages}{52--81}.
\newblock
\href{https://doi.org/10.1109/MCAS.2024.3349669}{doi:\nolinkurl{10.1109/MCAS.2024.3349669}}


\bibitem[Jouppi et~al\mbox{.}(2023)]%
        {tpuv4}
\bibfield{author}{\bibinfo{person}{Norm Jouppi}, \bibinfo{person}{George
  Kurian}, \bibinfo{person}{Sheng Li}, \bibinfo{person}{Peter Ma},
  \bibinfo{person}{Rahul Nagarajan}, \bibinfo{person}{Lifeng Nai},
  \bibinfo{person}{Nishant Patil}, \bibinfo{person}{Suvinay Subramanian},
  \bibinfo{person}{Andy Swing}, \bibinfo{person}{Brian Towles},
  \bibinfo{person}{Clifford Young}, \bibinfo{person}{Xiang Zhou},
  \bibinfo{person}{Zongwei Zhou}, {and} \bibinfo{person}{David~A Patterson}.}
  \bibinfo{year}{2023}\natexlab{}.
\newblock \showarticletitle{TPU v4: An Optically Reconfigurable Supercomputer
  for Machine Learning with Hardware Support for Embeddings}. In
  \bibinfo{booktitle}{\emph{Proceedings of the 50th Annual International
  Symposium on Computer Architecture}} (Orlando, FL, USA)
  \emph{(\bibinfo{series}{ISCA '23})}. \bibinfo{publisher}{Association for
  Computing Machinery}, \bibinfo{address}{New York, NY, USA}, Article
  \bibinfo{articleno}{82}, \bibinfo{numpages}{14}~pages.
\newblock
\showISBNx{9798400700958}
\href{https://doi.org/10.1145/3579371.3589350}{doi:\nolinkurl{10.1145/3579371.3589350}}


\bibitem[Kwon et~al\mbox{.}(2023)]%
        {vllm}
\bibfield{author}{\bibinfo{person}{Woosuk Kwon}, \bibinfo{person}{Zhuohan Li},
  \bibinfo{person}{Siyuan Zhuang}, \bibinfo{person}{Ying Sheng},
  \bibinfo{person}{Lianmin Zheng}, \bibinfo{person}{Cody~Hao Yu},
  \bibinfo{person}{Joseph~E. Gonzalez}, \bibinfo{person}{Hao Zhang}, {and}
  \bibinfo{person}{Ion Stoica}.} \bibinfo{year}{2023}\natexlab{}.
\newblock \bibinfo{title}{Efficient Memory Management for Large Language Model
  Serving with PagedAttention}.
\newblock
\showeprint[arxiv]{2309.06180}~[cs.LG]
\urldef\tempurl%
\url{https://arxiv.org/abs/2309.06180}
\showURL{%
\tempurl}


\bibitem[Liu et~al\mbox{.}(2023a)]%
        {failure1}
\bibfield{author}{\bibinfo{person}{Heting Liu}, \bibinfo{person}{Zhichao Li},
  \bibinfo{person}{Cheng Tan}, \bibinfo{person}{Rongqiu Yang},
  \bibinfo{person}{Guohong Cao}, \bibinfo{person}{Zherui Liu}, {and}
  \bibinfo{person}{Chuanxiong Guo}.} \bibinfo{year}{2023}\natexlab{a}.
\newblock \showarticletitle{Predicting GPU Failures With High Precision Under
  Deep Learning Workloads}. In \bibinfo{booktitle}{\emph{Proceedings of the
  16th ACM International Conference on Systems and Storage}}.
  \bibinfo{pages}{124--135}.
\newblock


\bibitem[Liu et~al\mbox{.}(2023b)]%
        {ringattn}
\bibfield{author}{\bibinfo{person}{Hao Liu}, \bibinfo{person}{Matei Zaharia},
  {and} \bibinfo{person}{Pieter Abbeel}.} \bibinfo{year}{2023}\natexlab{b}.
\newblock \bibinfo{title}{Ring Attention with Blockwise Transformers for
  Near-Infinite Context}.
\newblock
\showeprint[arxiv]{2310.01889}~[cs.CL]
\urldef\tempurl%
\url{https://arxiv.org/abs/2310.01889}
\showURL{%
\tempurl}


\bibitem[Liu and Wong(2024)]%
        {retical}
\bibfield{author}{\bibinfo{person}{Mark Liu} {and}
  \bibinfo{person}{H.-S.~Philip Wong}.} \bibinfo{year}{2024}\natexlab{}.
\newblock \showarticletitle{How We’ll Reach a 1 Trillion Transistor GPU}.
\newblock \bibinfo{journal}{\emph{IEEE Spectrum}} (\bibinfo{date}{March}
  \bibinfo{year}{2024}).
\newblock
\urldef\tempurl%
\url{https://spectrum.ieee.org/trillion-transistor-gpu}
\showURL{%
\tempurl}
\newblock
\shownote{Accessed: 2025-01-15}.


\bibitem[Loh et~al\mbox{.}(2021)]%
        {chiplet1}
\bibfield{author}{\bibinfo{person}{Gabriel~H Loh}, \bibinfo{person}{Samuel
  Naffziger}, {and} \bibinfo{person}{Kevin Lepak}.}
  \bibinfo{year}{2021}\natexlab{}.
\newblock \showarticletitle{Understanding chiplets today to anticipate future
  integration opportunities and limits}. In \bibinfo{booktitle}{\emph{2021
  Design, Automation \& Test in Europe Conference \& Exhibition (DATE)}}. IEEE,
  \bibinfo{pages}{142--145}.
\newblock


\bibitem[Maruf and Chowdhury(2023)]%
        {disaggmem}
\bibfield{author}{\bibinfo{person}{Hasan~Al Maruf} {and}
  \bibinfo{person}{Mosharaf Chowdhury}.} \bibinfo{year}{2023}\natexlab{}.
\newblock \bibinfo{title}{Memory Disaggregation: Advances and Open Challenges}.
\newblock
\showeprint[arxiv]{2305.03943}~[cs.DC]
\urldef\tempurl%
\url{https://arxiv.org/abs/2305.03943}
\showURL{%
\tempurl}


\bibitem[Meta({[n.\,d.]})]%
        {llama_train}
\bibfield{author}{\bibinfo{person}{Meta}.} \bibinfo{year}{[n.\,d.]}\natexlab{}.
\newblock \bibinfo{title}{{I}ntroducing {L}lama 3.1: {O}ur most capable models
  to date}.
\newblock
  \bibinfo{howpublished}{\url{https://ai.meta.com/blog/meta-llama-3-1/}}.
\newblock
\newblock
\shownote{[Accessed 15-04-2025]}.


\bibitem[Meta(2025)]%
        {llama}
\bibfield{author}{\bibinfo{person}{Meta}.} \bibinfo{year}{2025}\natexlab{}.
\newblock \bibinfo{title}{Meta Llama on Hugging Face}.
\newblock \bibinfo{howpublished}{\url{https://huggingface.co/meta-llama}}.
\newblock
\newblock
\shownote{Accessed: 2025-01-13}.


\bibitem[Miao et~al\mbox{.}(2024)]%
        {faulttol2}
\bibfield{author}{\bibinfo{person}{Xupeng Miao}, \bibinfo{person}{Chunan Shi},
  \bibinfo{person}{Jiangfei Duan}, \bibinfo{person}{Xiaoli Xi},
  \bibinfo{person}{Dahua Lin}, \bibinfo{person}{Bin Cui}, {and}
  \bibinfo{person}{Zhihao Jia}.} \bibinfo{year}{2024}\natexlab{}.
\newblock \showarticletitle{SpotServe: Serving Generative Large Language Models
  on Preemptible Instances}. In \bibinfo{booktitle}{\emph{Proceedings of the
  29th ACM International Conference on Architectural Support for Programming
  Languages and Operating Systems, Volume 2}} (La Jolla, CA, USA)
  \emph{(\bibinfo{series}{ASPLOS '24})}. \bibinfo{publisher}{Association for
  Computing Machinery}, \bibinfo{address}{New York, NY, USA},
  \bibinfo{pages}{1112–1127}.
\newblock
\showISBNx{9798400703850}
\href{https://doi.org/10.1145/3620665.3640411}{doi:\nolinkurl{10.1145/3620665.3640411}}


\bibitem[Minaee et~al\mbox{.}(2024)]%
        {modelsurvey}
\bibfield{author}{\bibinfo{person}{Shervin Minaee}, \bibinfo{person}{Tomas
  Mikolov}, \bibinfo{person}{Narjes Nikzad}, \bibinfo{person}{Meysam
  Chenaghlu}, \bibinfo{person}{Richard Socher}, \bibinfo{person}{Xavier
  Amatriain}, {and} \bibinfo{person}{Jianfeng Gao}.}
  \bibinfo{year}{2024}\natexlab{}.
\newblock \bibinfo{title}{Large Language Models: A Survey}.
\newblock
\showeprint[arxiv]{2402.06196}~[cs.CL]
\urldef\tempurl%
\url{https://arxiv.org/abs/2402.06196}
\showURL{%
\tempurl}


\bibitem[Minkenberg et~al\mbox{.}(2021)]%
        {cpo1}
\bibfield{author}{\bibinfo{person}{Cyriel Minkenberg},
  \bibinfo{person}{Rajagopal Krishnaswamy}, \bibinfo{person}{Aaron Zilkie},
  {and} \bibinfo{person}{David Nelson}.} \bibinfo{year}{2021}\natexlab{}.
\newblock \showarticletitle{Co-packaged datacenter optics: Opportunities and
  challenges}.
\newblock \bibinfo{journal}{\emph{IET optoelectronics}} \bibinfo{volume}{15},
  \bibinfo{number}{2} (\bibinfo{year}{2021}), \bibinfo{pages}{77--91}.
\newblock


\bibitem[{Moore Elite}({[n.\,d.]})]%
        {yield_calc}
\bibfield{author}{\bibinfo{person}{{Moore Elite}}.}
  \bibinfo{year}{[n.\,d.]}\natexlab{}.
\newblock \bibinfo{title}{Die Yield Calculator}.
\newblock
  \bibinfo{howpublished}{\url{http://cloud.mooreelite.com/tools/die-yield-calculator/index.html}}.
\newblock
\newblock
\shownote{Accessed: 2025-01-15}.


\bibitem[Morales(2024)]%
        {blackwelldelayed}
\bibfield{author}{\bibinfo{person}{Jowi Morales}.}
  \bibinfo{year}{2024}\natexlab{}.
\newblock \showarticletitle{Nvidia Blackwell GPUs Allegedly Delayed Due to
  Design Flaws}.
\newblock \bibinfo{journal}{\emph{Tom's Hardware}} (\bibinfo{year}{2024}).
\newblock
\urldef\tempurl%
\url{{https://www.tomshardware.com/pc-components/gpus/nvidia-blackwell-gpus-allegedly-delayed-due-to-design-flaws}}
\showURL{%
\tempurl}
\newblock
\shownote{Accessed: 2025-01-10}.


\bibitem[NVIDIA({[n.\,d.]})]%
        {nvidia_cpo_gtc}
\bibfield{author}{\bibinfo{person}{NVIDIA}.}
  \bibinfo{year}{[n.\,d.]}\natexlab{}.
\newblock \bibinfo{title}{{N}{V}{I}{D}{I}{A} {A}nnounces {S}pectrum-{X}
  {P}hotonics, {C}o-{P}ackaged {O}ptics {N}etworking {S}witches to {S}cale
  {A}{I} {F}actories to {M}illions of {G}{P}{U}s}.
\newblock
  \bibinfo{howpublished}{\url{https://nvidianews.nvidia.com/news/nvidia-spectrum-x-co-packaged-optics-networking-switches-ai-factories}}.
\newblock
\newblock
\shownote{[Accessed 15-04-2025]}.


\bibitem[{NVIDIA}(2025)]%
        {digits}
\bibfield{author}{\bibinfo{person}{{NVIDIA}}.} \bibinfo{year}{2025}\natexlab{}.
\newblock \bibinfo{title}{NVIDIA Project DIGITS}.
\newblock
  \bibinfo{howpublished}{\url{https://www.nvidia.com/en-us/project-digits/}}.
\newblock
\newblock
\shownote{Accessed: 2025-01-14}.


\bibitem[Patel et~al\mbox{.}(2024)]%
        {splitwise}
\bibfield{author}{\bibinfo{person}{Pratyush Patel}, \bibinfo{person}{Esha
  Choukse}, \bibinfo{person}{Chaojie Zhang}, \bibinfo{person}{Aashaka Shah},
  \bibinfo{person}{{\'I}{\~n}igo Goiri}, \bibinfo{person}{Saeed Maleki}, {and}
  \bibinfo{person}{Ricardo Bianchini}.} \bibinfo{year}{2024}\natexlab{}.
\newblock \showarticletitle{Splitwise: Efficient generative llm inference using
  phase splitting}. In \bibinfo{booktitle}{\emph{2024 ACM/IEEE 51st Annual
  International Symposium on Computer Architecture (ISCA)}}. IEEE,
  \bibinfo{pages}{118--132}.
\newblock


\bibitem[Piga et~al\mbox{.}(2024)]%
        {dvfs_meta}
\bibfield{author}{\bibinfo{person}{Leonardo Piga}, \bibinfo{person}{Iyswarya
  Narayanan}, \bibinfo{person}{Aditya Sundarrajan}, \bibinfo{person}{Matt
  Skach}, \bibinfo{person}{Qingyuan Deng}, \bibinfo{person}{Biswadip Maity},
  \bibinfo{person}{Manoj Chakkaravarthy}, \bibinfo{person}{Alison Huang},
  \bibinfo{person}{Abhishek Dhanotia}, {and} \bibinfo{person}{Parth Malani}.}
  \bibinfo{year}{2024}\natexlab{}.
\newblock \showarticletitle{Expanding Datacenter Capacity with DVFS Boosting: A
  safe and scalable deployment experience}. In
  \bibinfo{booktitle}{\emph{Proceedings of the 29th ACM International
  Conference on Architectural Support for Programming Languages and Operating
  Systems, Volume 1}} (La Jolla, CA, USA) \emph{(\bibinfo{series}{ASPLOS
  '24})}. \bibinfo{publisher}{Association for Computing Machinery},
  \bibinfo{address}{New York, NY, USA}, \bibinfo{pages}{150–165}.
\newblock
\showISBNx{9798400703720}
\href{https://doi.org/10.1145/3617232.3624853}{doi:\nolinkurl{10.1145/3617232.3624853}}


\bibitem[Qiu et~al\mbox{.}(2024)]%
        {userve}
\bibfield{author}{\bibinfo{person}{Haoran Qiu}, \bibinfo{person}{Weichao Mao},
  \bibinfo{person}{Archit Patke}, \bibinfo{person}{Shengkun Cui},
  \bibinfo{person}{Saurabh Jha}, \bibinfo{person}{Chen Wang},
  \bibinfo{person}{Hubertus Franke}, \bibinfo{person}{Zbigniew Kalbarczyk},
  \bibinfo{person}{Tamer Ba{\c{s}}ar}, {and} \bibinfo{person}{Ravishankar~K
  Iyer}.} \bibinfo{year}{2024}\natexlab{}.
\newblock \showarticletitle{Power-aware Deep Learning Model Serving with
  $\{$$\mu$-Serve$\}$}. In \bibinfo{booktitle}{\emph{2024 USENIX Annual
  Technical Conference (USENIX ATC 24)}}. \bibinfo{pages}{75--93}.
\newblock


\bibitem[Shah et~al\mbox{.}(2024)]%
        {fav3}
\bibfield{author}{\bibinfo{person}{Jay Shah}, \bibinfo{person}{Ganesh
  Bikshandi}, \bibinfo{person}{Ying Zhang}, \bibinfo{person}{Vijay Thakkar},
  \bibinfo{person}{Pradeep Ramani}, {and} \bibinfo{person}{Tri Dao}.}
  \bibinfo{year}{2024}\natexlab{}.
\newblock \showarticletitle{Flashattention-3: Fast and accurate attention with
  asynchrony and low-precision}.
\newblock \bibinfo{journal}{\emph{arXiv preprint arXiv:2407.08608}}
  (\bibinfo{year}{2024}).
\newblock


\bibitem[Shoeybi et~al\mbox{.}(2020)]%
        {megatron}
\bibfield{author}{\bibinfo{person}{Mohammad Shoeybi}, \bibinfo{person}{Mostofa
  Patwary}, \bibinfo{person}{Raul Puri}, \bibinfo{person}{Patrick LeGresley},
  \bibinfo{person}{Jared Casper}, {and} \bibinfo{person}{Bryan Catanzaro}.}
  \bibinfo{year}{2020}\natexlab{}.
\newblock \bibinfo{title}{Megatron-LM: Training Multi-Billion Parameter
  Language Models Using Model Parallelism}.
\newblock
\showeprint[arxiv]{1909.08053}~[cs.CL]
\urldef\tempurl%
\url{https://arxiv.org/abs/1909.08053}
\showURL{%
\tempurl}


\bibitem[Song et~al\mbox{.}(2024)]%
        {powerinfer}
\bibfield{author}{\bibinfo{person}{Yixin Song}, \bibinfo{person}{Zeyu Mi},
  \bibinfo{person}{Haotong Xie}, {and} \bibinfo{person}{Haibo Chen}.}
  \bibinfo{year}{2024}\natexlab{}.
\newblock \showarticletitle{Powerinfer: Fast large language model serving with
  a consumer-grade gpu}. In \bibinfo{booktitle}{\emph{Proceedings of the ACM
  SIGOPS 30th Symposium on Operating Systems Principles}}.
  \bibinfo{pages}{590--606}.
\newblock


\bibitem[Stojkovic et~al\mbox{.}(2024a)]%
        {greener}
\bibfield{author}{\bibinfo{person}{Jovan Stojkovic}, \bibinfo{person}{Esha
  Choukse}, \bibinfo{person}{Chaojie Zhang}, \bibinfo{person}{Inigo Goiri},
  {and} \bibinfo{person}{Josep Torrellas}.} \bibinfo{year}{2024}\natexlab{a}.
\newblock \showarticletitle{Towards Greener LLMs: Bringing Energy-Efficiency to
  the Forefront of LLM Inference}.
\newblock \bibinfo{journal}{\emph{arXiv preprint arXiv:2403.20306}}
  (\bibinfo{year}{2024}).
\newblock


\bibitem[Stojkovic et~al\mbox{.}(2024b)]%
        {overclock_ms}
\bibfield{author}{\bibinfo{person}{Jovan Stojkovic}, \bibinfo{person}{Pulkit~A.
  Misra}, \bibinfo{person}{Íñigo Goiri}, \bibinfo{person}{Sam Whitlock},
  \bibinfo{person}{Esha Choukse}, \bibinfo{person}{Mayukh Das},
  \bibinfo{person}{Chetan Bansal}, \bibinfo{person}{Jason Lee},
  \bibinfo{person}{Zoey Sun}, \bibinfo{person}{Haoran Qiu},
  \bibinfo{person}{Reed Zimmermann}, \bibinfo{person}{Savyasachi Samal},
  \bibinfo{person}{Brijesh Warrier}, \bibinfo{person}{Ashish Raniwala}, {and}
  \bibinfo{person}{Ricardo Bianchini}.} \bibinfo{year}{2024}\natexlab{b}.
\newblock \showarticletitle{SmartOClock: Workload- and Risk-Aware Overclocking
  in the Cloud}. In \bibinfo{booktitle}{\emph{2024 ACM/IEEE 51st Annual
  International Symposium on Computer Architecture (ISCA)}}.
  \bibinfo{pages}{437--451}.
\newblock
\href{https://doi.org/10.1109/ISCA59077.2024.00040}{doi:\nolinkurl{10.1109/ISCA59077.2024.00040}}


\bibitem[Strati et~al\mbox{.}(2024)]%
        {faulttol1}
\bibfield{author}{\bibinfo{person}{Foteini Strati}, \bibinfo{person}{Sara
  Mcallister}, \bibinfo{person}{Amar Phanishayee}, \bibinfo{person}{Jakub
  Tarnawski}, {and} \bibinfo{person}{Ana Klimovic}.}
  \bibinfo{year}{2024}\natexlab{}.
\newblock \showarticletitle{DéjàVu: {KV}-cache Streaming for Fast,
  Fault-tolerant Generative {LLM} Serving}. In
  \bibinfo{booktitle}{\emph{Proceedings of the 41st International Conference on
  Machine Learning}} \emph{(\bibinfo{series}{Proceedings of Machine Learning
  Research}, Vol.~\bibinfo{volume}{235})},
  \bibfield{editor}{\bibinfo{person}{Ruslan Salakhutdinov},
  \bibinfo{person}{Zico Kolter}, \bibinfo{person}{Katherine Heller},
  \bibinfo{person}{Adrian Weller}, \bibinfo{person}{Nuria Oliver},
  \bibinfo{person}{Jonathan Scarlett}, {and} \bibinfo{person}{Felix
  Berkenkamp}} (Eds.). \bibinfo{publisher}{PMLR},
  \bibinfo{pages}{46745--46771}.
\newblock
\urldef\tempurl%
\url{https://proceedings.mlr.press/v235/strati24a.html}
\showURL{%
\tempurl}


\bibitem[Sun et~al\mbox{.}(2024)]%
        {llumnix}
\bibfield{author}{\bibinfo{person}{Biao Sun}, \bibinfo{person}{Ziming Huang},
  \bibinfo{person}{Hanyu Zhao}, \bibinfo{person}{Wencong Xiao},
  \bibinfo{person}{Xinyi Zhang}, \bibinfo{person}{Yong Li}, {and}
  \bibinfo{person}{Wei Lin}.} \bibinfo{year}{2024}\natexlab{}.
\newblock \showarticletitle{Llumnix: Dynamic Scheduling for Large Language
  Model Serving}.
\newblock \bibinfo{journal}{\emph{arXiv preprint arXiv:2406.03243}}
  (\bibinfo{year}{2024}).
\newblock


\bibitem[Tan et~al\mbox{.}(2023)]%
        {cpo2}
\bibfield{author}{\bibinfo{person}{Min Tan}, \bibinfo{person}{Jiang Xu},
  \bibinfo{person}{Siyang Liu}, \bibinfo{person}{Junbo Feng},
  \bibinfo{person}{Hua Zhang}, \bibinfo{person}{Chaonan Yao},
  \bibinfo{person}{Shixi Chen}, \bibinfo{person}{Hangyu Guo},
  \bibinfo{person}{Gengshi Han}, \bibinfo{person}{Zhanhao Wen},
  {et~al\mbox{.}}} \bibinfo{year}{2023}\natexlab{}.
\newblock \showarticletitle{Co-packaged optics (CPO): status, challenges, and
  solutions}.
\newblock \bibinfo{journal}{\emph{Frontiers of Optoelectronics}}
  \bibinfo{volume}{16}, \bibinfo{number}{1} (\bibinfo{year}{2023}),
  \bibinfo{pages}{1}.
\newblock


\bibitem[Tang and Xie(2022)]%
        {packagingcost}
\bibfield{author}{\bibinfo{person}{Tianqi Tang} {and} \bibinfo{person}{Yuan
  Xie}.} \bibinfo{year}{2022}\natexlab{}.
\newblock \bibinfo{title}{Cost-Aware Exploration for Chiplet-Based Architecture
  with Advanced Packaging Technologies}.
\newblock
\showeprint[arxiv]{2206.07308}~[cs.AR]
\urldef\tempurl%
\url{https://arxiv.org/abs/2206.07308}
\showURL{%
\tempurl}


\bibitem[Team(2025)]%
        {blackwelloverheat}
\bibfield{author}{\bibinfo{person}{DW Team}.} \bibinfo{year}{2025}\natexlab{}.
\newblock \showarticletitle{Nvidia faces order delays as Blackwell chips
  overheat}.
\newblock \bibinfo{journal}{\emph{digwatch}} (\bibinfo{year}{2025}).
\newblock
\urldef\tempurl%
\url{{https://dig.watch/updates/nvidia-faces-order-delays-as-blackwell-chips-overheat}}
\showURL{%
\tempurl}
\newblock
\shownote{Accessed: 2025-01-14}.


\bibitem[Teets(1996)]%
        {yield2}
\bibfield{author}{\bibinfo{person}{D. Teets}.} \bibinfo{year}{1996}\natexlab{}.
\newblock \showarticletitle{A model for radial yield degradation as a function
  of chip size}.
\newblock \bibinfo{journal}{\emph{IEEE Transactions on Semiconductor
  Manufacturing}} \bibinfo{volume}{9}, \bibinfo{number}{3}
  (\bibinfo{year}{1996}), \bibinfo{pages}{467--471}.
\newblock
\href{https://doi.org/10.1109/66.536118}{doi:\nolinkurl{10.1109/66.536118}}


\bibitem[{The Mac Observer}(2025)]%
        {neuralengine}
\bibfield{author}{\bibinfo{person}{{The Mac Observer}}.}
  \bibinfo{year}{2025}\natexlab{}.
\newblock \bibinfo{title}{What is Apple Neural Engine?}
\newblock
\urldef\tempurl%
\url{https://www.macobserver.com/tips/what-is-apple-neural-engine/}
\showURL{%
\tempurl}
\newblock
\shownote{Accessed: 2025-01-14}.


\bibitem[Tirumala and Wong(2024)]%
        {blackwell}
\bibfield{author}{\bibinfo{person}{Ajay Tirumala} {and}
  \bibinfo{person}{Raymond Wong}.} \bibinfo{year}{2024}\natexlab{}.
\newblock \showarticletitle{NVIDIA Blackwell Platform: Advancing Generative AI
  and Accelerated Computing}. In \bibinfo{booktitle}{\emph{2024 IEEE Hot Chips
  36 Symposium (HCS)}}. IEEE Computer Society, \bibinfo{pages}{1--33}.
\newblock


\bibitem[Vaswani et~al\mbox{.}(2023)]%
        {attentionneed}
\bibfield{author}{\bibinfo{person}{Ashish Vaswani}, \bibinfo{person}{Noam
  Shazeer}, \bibinfo{person}{Niki Parmar}, \bibinfo{person}{Jakob Uszkoreit},
  \bibinfo{person}{Llion Jones}, \bibinfo{person}{Aidan~N. Gomez},
  \bibinfo{person}{Lukasz Kaiser}, {and} \bibinfo{person}{Illia Polosukhin}.}
  \bibinfo{year}{2023}\natexlab{}.
\newblock \bibinfo{title}{Attention Is All You Need}.
\newblock
\showeprint[arxiv]{1706.03762}~[cs.CL]
\urldef\tempurl%
\url{https://arxiv.org/abs/1706.03762}
\showURL{%
\tempurl}


\bibitem[Williams et~al\mbox{.}(2009)]%
        {roofline}
\bibfield{author}{\bibinfo{person}{Samuel Williams}, \bibinfo{person}{Andrew
  Waterman}, {and} \bibinfo{person}{David Patterson}.}
  \bibinfo{year}{2009}\natexlab{}.
\newblock \showarticletitle{Roofline: an insightful visual performance model
  for multicore architectures}.
\newblock \bibinfo{journal}{\emph{Commun. ACM}} \bibinfo{volume}{52},
  \bibinfo{number}{4} (\bibinfo{date}{April} \bibinfo{year}{2009}),
  \bibinfo{pages}{65–76}.
\newblock
\showISSN{0001-0782}
\href{https://doi.org/10.1145/1498765.1498785}{doi:\nolinkurl{10.1145/1498765.1498785}}


\bibitem[Xu et~al\mbox{.}(2024b)]%
        {mllmnpu}
\bibfield{author}{\bibinfo{person}{Daliang Xu}, \bibinfo{person}{Hao Zhang},
  \bibinfo{person}{Liming Yang}, \bibinfo{person}{Ruiqi Liu},
  \bibinfo{person}{Gang Huang}, \bibinfo{person}{Mengwei Xu}, {and}
  \bibinfo{person}{Xuanzhe Liu}.} \bibinfo{year}{2024}\natexlab{b}.
\newblock \showarticletitle{Empowering 1000 tokens/second on-device llm
  prefilling with mllm-npu}.
\newblock \bibinfo{journal}{\emph{arXiv preprint arXiv:2407.05858}}
  (\bibinfo{year}{2024}).
\newblock


\bibitem[Xu et~al\mbox{.}(2024c)]%
        {infnpu}
\bibfield{author}{\bibinfo{person}{Daliang Xu}, \bibinfo{person}{Hao Zhang},
  \bibinfo{person}{Liming Yang}, \bibinfo{person}{Ruiqi Liu},
  \bibinfo{person}{Gang Huang}, \bibinfo{person}{Mengwei Xu}, {and}
  \bibinfo{person}{Xuanzhe Liu}.} \bibinfo{year}{2024}\natexlab{c}.
\newblock \bibinfo{title}{Fast On-device LLM Inference with NPUs}.
\newblock
\showeprint[arxiv]{2407.05858}~[cs.AI]
\urldef\tempurl%
\url{https://arxiv.org/abs/2407.05858}
\showURL{%
\tempurl}


\bibitem[Xu et~al\mbox{.}(2024a)]%
        {ondevice}
\bibfield{author}{\bibinfo{person}{Jiajun Xu}, \bibinfo{person}{Zhiyuan Li},
  \bibinfo{person}{Wei Chen}, \bibinfo{person}{Qun Wang}, \bibinfo{person}{Xin
  Gao}, \bibinfo{person}{Qi Cai}, {and} \bibinfo{person}{Ziyuan Ling}.}
  \bibinfo{year}{2024}\natexlab{a}.
\newblock \bibinfo{title}{On-Device Language Models: A Comprehensive Review}.
\newblock
\showeprint[arxiv]{2409.00088}~[cs.CL]
\urldef\tempurl%
\url{https://arxiv.org/abs/2409.00088}
\showURL{%
\tempurl}


\bibitem[Xu et~al\mbox{.}(2021)]%
        {paralleltraining}
\bibfield{author}{\bibinfo{person}{Weizheng Xu}, \bibinfo{person}{Youtao
  Zhang}, {and} \bibinfo{person}{Xulong Tang}.}
  \bibinfo{year}{2021}\natexlab{}.
\newblock \showarticletitle{Parallelizing DNN Training on GPUs: Challenges and
  Opportunities}. In \bibinfo{booktitle}{\emph{Companion Proceedings of the Web
  Conference 2021}} (Ljubljana, Slovenia) \emph{(\bibinfo{series}{WWW '21})}.
  \bibinfo{publisher}{Association for Computing Machinery},
  \bibinfo{address}{New York, NY, USA}, \bibinfo{pages}{174–178}.
\newblock
\showISBNx{9781450383134}
\href{https://doi.org/10.1145/3442442.3452055}{doi:\nolinkurl{10.1145/3442442.3452055}}


\bibitem[Zhang and Jia(2022)]%
        {opticalreach}
\bibfield{author}{\bibinfo{person}{Junwen Zhang} {and}
  \bibinfo{person}{Zhensheng Jia}.} \bibinfo{year}{2022}\natexlab{}.
\newblock \showarticletitle{Coherent Passive Optical Networks for
  100G/$\lambda$-and-Beyond Fiber Access: Recent Progress and Outlook}.
\newblock \bibinfo{journal}{\emph{IEEE Network}} \bibinfo{volume}{36},
  \bibinfo{number}{2} (\bibinfo{year}{2022}), \bibinfo{pages}{116--123}.
\newblock
\href{https://doi.org/10.1109/MNET.005.2100604}{doi:\nolinkurl{10.1109/MNET.005.2100604}}


\bibitem[Zhong et~al\mbox{.}(2024)]%
        {distserve}
\bibfield{author}{\bibinfo{person}{Yinmin Zhong}, \bibinfo{person}{Shengyu
  Liu}, \bibinfo{person}{Junda Chen}, \bibinfo{person}{Jianbo Hu},
  \bibinfo{person}{Yibo Zhu}, \bibinfo{person}{Xuanzhe Liu},
  \bibinfo{person}{Xin Jin}, {and} \bibinfo{person}{Hao Zhang}.}
  \bibinfo{year}{2024}\natexlab{}.
\newblock \bibinfo{title}{DistServe: Disaggregating Prefill and Decoding for
  Goodput-optimized Large Language Model Serving}.
\newblock
\showeprint[arxiv]{2401.09670}~[cs.DC]
\urldef\tempurl%
\url{https://arxiv.org/abs/2401.09670}
\showURL{%
\tempurl}


\end{thebibliography}

\end{document}